\documentclass[doublecol]{epl2}

\title{Fermion localization on branes with generalized dynamics}
\shorttitle{Fermion localization on branes with generalized dynamics} 

\author{L. B. Castro\inst{1}\thanks{E.mail: \email{luis.castro@pgfsc.ufsc.br}} %
        \and L. E. Arroyo Meza\inst{2}\thanks{E.mail: \email{luisarroyo@feg.unesp.br}}}
\shortauthor{L. B. Castro and L. E. Arroyo Meza}

\institute{
  \inst{1} Departamento de F\'{\i}sica CFM, Universidade Federal de Santa Catarina, %
  Florian\'{o}polis, SC CP. 476, CEP 88.040-900, Brazil.\\
  \inst{2} Departamento de F\'{\i}sica e Qu\'{\i}mica, Universidade Estadual Paulista, %
  12516-410 Guaratinguet\'{a}, S\~ao Paulo, Brazil.
}

\abstract{
In this letter we consider a specific model of braneworld with nonstandard dynamics
di\-ffu\-sed in the literature, specifically we focus our attention on the matter energy density, the
energy of system, the Ricci scalar and the thin brane limit. As the model is classically stable and
capable of localize gravity, as a natural extension we address the issue of fermion localization of
fermions on a thick brane constructed out from one scalar field with nonstandard kinetic terms coupled
with gravity. The contribution of the nonstandard kinetic terms in the problem of fermion localization
is analyzed. It is found that the simplest Yukawa coupling $\eta\bar{\Psi}\phi\Psi$ support the
lo\-ca\-li\-za\-tion of fermions on the thick brane. It is shown that the zero mode for left-handed
can be localized on the thick brane depending on the values for the coupling constant $\eta$.
}

\begin{document}

\maketitle

\section{Introduction}
In the last decade, the braneworld scenario has attracted a lot of interests for it gives an
effec\-tive way to solve the hierarchy problem by introdu\-cing two 3-branes which are embedded
in a 5-dimensional anti-de Sitter ($AdS_{5}$) space-time~\cite{rs1}. Parallel researchs on noncompact extra dimensions focused on trapping ma\-tter and model of the Universe as a 3-shell expanding in 5-dimensional space-time~\cite{go} have also attracted interest. As another attractive property, the Newtonian law of gravity with a correction is also given in this braneworld
scenario~\cite{rs2}. In the Randall-Sundrum (RS) model~\cite{rs1}, we can further add scalar fields~\cite{gol} with usual dynamics and allow them to interact with gravity in the standard way. In this scenario, the smooth character of the solutions generate thick brane with a diversity of structures~\cite{sken,wol,grem,ba}. In the braneworld scenarios, an important issue is how
gravity and different observable matter fields of the Standard Model of particle physics are
localized on the brane. It has been shown that, in the RS model in 5-dimensional
space-time, graviton and spin 0 field can be localized on a brane with positive tension%
~\cite{rs2,ran,bajc}. Moreover, spin 1/2 and 3/2 can be localized on a negative-tension brane%
~\cite{bajc}. The localization problem of spin-1/2 fermions on thick branes is interesting and
important~\cite{ran,bajc,gross,dub,mou,rin,mel,sing,sing2,gib,xia,casa,yo,kk}.
In order to achieve localization of fermions on a brane with positive tension, it seems that
additional interactions except the gravitational interaction must be including in the bulk.

On the other hand, the first recent observations~\cite{perl} have led us with the
intriguing fact that the Universe is presently undertaking accelerated expansion. These information
directly contributed to establish some important advances in cosmology, one of them being
the presence of dark energy. The presence of dark energy has opened some distinct routes
of investigations. In recent years, there appeared some interesting models with noncanonical
dynamics with focus on early time inflation or dark energy~\cite{ba3,chiba,arme,arme2}, as
for instance, the so-called k-fields, first introduced in the context of inflation~\cite{arme2}
and the k-essence models, suggested to solve the cosmic coincidence problem~\cite{arme,arme2,arme3}.
The interaction between dark energy and fermion fields has already a precedent in the
cosmology context~\cite{darke}. In the context of braneworld scenario, the effect of general
brane kinetic terms for bulks scalars, fermions and gauge bosons in theories with and without
supersymmetry has already been analyzed in~\cite{agui}. We believe that the conditions for
obtaining zero modes in braneworld models of scalar fields with generalized dynamics deserves to
be more explored.

The purpose of the present letter is to analyze the contribution of the nonstandard kinetic terms of a brane model in the problem of fermion localization. To achieve our
goal, the braneworld model with nonstandard dynamics $\mathcal{L}=K\left(X\right)-V(\phi)$, where $K\left(X\right)=X+\alpha|X|X$ (type I model in~\cite{ba3}) is considered. We will focus our attention mainly on the matter energy, the energy of system, the Ricci scalar and the thin brane
limit. As the model is classically stable and capable of localize gravity, additionally we address
ourselves to the issue of fermion localization on a thick brane constructed out from one scalar
field with nonstandard kinetic terms coupled to gravity. Following the same procedure of%
~\cite{ba3}, we use the analytical expressions for small $\alpha$ and investigate the contribution
of this nonstandard kinetic terms in the problem of fermion localization. We find that the simplest
Yukawa coupling $\eta\bar{\Psi}\phi\Psi$, where $\eta$ is the coupling constant, which allows
left-handed to posses a localized zero mode on the thick brane under some conditions on the value
for the coupling constant $\eta$. For $\alpha$, which is not necessarily small, we can not get
any expressions for the solution of this model, for this case the numerical study is used. The
numerical results bear out our results for small $\alpha$. The organization of this paper is as
follows: in Sec. II, we give a brief review of the model with generalized dynamics developed
by Bazeia and collaborators~\cite{ba3}. In Sec. III, we study the localization of spin-1/2
fermion for this model and we also analyze the essential conditions for the localization with the
simplest Yukawa coupling. Finally, our conclusions are presented in Sec. IV.

\section{Review of systems with generalized dynamics}
The action for this kind of system is described by~\cite{ba3}
\begin{equation}\label{e1}
S= \int d^{5}x \sqrt{|\,g|}\,\left[-\frac{1}{4}R + \mathcal{L}(\phi,X) \right],
\end{equation}
\noindent where $g\equiv \mathrm{Det}(g_{ab})$ and $X=\frac{1}{2}\,\nabla^{a}\phi\nabla_{a}\phi$.
The line element of the five-dimensional space-time can be written as
\begin{equation}\label{metric}
    ds^{2}=g_{ab}dx^{a}dx^{b}=\mathrm{e}^{2A(y)}\eta_{\mu\nu}dx^{\mu}dx^{\nu}-dy^{2},
\end{equation}
\noindent where we are using the five-dimensional Newton constat $4\pi G^{(5)}=1$, $y=x^{4}$
is the extra dimension (the Latin indices run from $0$ to $4$), $\eta_{\mu\nu}$ is the Minkowski
metric with signature $(+,-,-,-)$ and $\mathrm{e}^{2A}$ is the so-called warp factor (the Greek
indices run from $0$ to $3$). We suppose that $A=A(y)$ and $\phi=\phi(y)$.

One can determine the static equations of motion for the above system
are of the form
\begin{equation}\label{em1}
    \left(\mathcal{L}_{X}+2X\mathcal{L}_{XX}\right)\phi^{\prime\prime}-
    \left( 2X\mathcal{L}_{X\phi}-\mathcal{L}_{\phi} \right)= -
    4\mathcal{L}_{X}A^{\prime}\phi^{\prime},
\end{equation}
\begin{equation}\label{em2}
    A^{\prime\prime}+2A^{\prime}\,^{2}=\frac{2}{3}\mathcal{L},
\end{equation}
\begin{equation}\label{em3}
    A^{\prime}\,^{2}=\frac{1}{3}\left( \mathcal{L}-2X\mathcal{L}_{X} \right),
\end{equation}
\noindent where prime stands for derivate with respect to $y$, $\mathcal{L}_{X}=\partial\mathcal{L}%
/\partial X$ and $\mathcal{L}_{\phi}=\partial\mathcal{L}/\partial\phi$.
\noindent Furthermore, the matter energy density is given by
\begin{equation}\label{de}
    \rho(y)=-\mathrm{e}^{2A(y)}\mathcal{L},
\end{equation}
\noindent and the scalar curvature (or Ricci scalar) is given by
\begin{equation}\label{ricci}
    R=-4(5A^{\prime}\,^{2}+2A^{\prime\prime}).
\end{equation}

The Lagrangian density $\mathcal{L}(\phi,X)$ has the form
\begin{equation}\label{dlg}
    \mathcal{L}=K(X)-V(\phi),
\end{equation}
\noindent where $K(X)$ and $V(\phi)$ are nonstandard kinetic term and
potential respectively.
\noindent For this case, from eqs.~(\ref{em1}),~(\ref{em2}) and~(\ref{em3})
the equations of motion can be expressed as
\begin{equation}\label{em1a}
    \left(K^{\prime}+2XK^{\prime\prime}\right)\phi^{\prime\prime}-V_{\phi}=
    -4K^{\prime}A^{\prime}\phi^{\prime},
\end{equation}
\begin{equation}\label{em2a}
     A^{\prime\prime}+2A^{\prime}\,^{2}=\frac{2}{3}(K-V),
\end{equation}
\begin{equation}\label{em3a}
    A^{\prime}\,^{2}=\frac{1}{3}\left( K-V-2XK^{\prime} \right).
\end{equation}
\noindent These equations are the static equations of motion of a system with nonstandard
dynamics. In~\cite{ba3}, the authors present two explicit models for $K(X)$, here we review
one of these models.

\subsection{The model: $K(X)=X+\alpha|X|X$}
\label{sec:2}
This model is also considered in~\cite{adam}, where $\alpha$ is a real non-nega%
tive parameter and $X=-\frac{1}{2}\,\phi^{\prime}\,^{2}$. If $\alpha=0$ the
standard scenario is restored. For this model, the equations of motion are
\begin{equation}\label{em1b}
    \phi^{\prime\prime}+4A^{\prime}\phi^{\prime}-V_{\phi}= -\alpha\left(3\phi^{\prime\prime}
    +4\phi^{\prime}A^{\prime}\right)\phi^{\prime}\,^{2},
\end{equation}
\begin{equation}\label{em2b}
    A^{\prime\prime}+2A^{\prime}\,^{2}=-\frac{1}{3}\,\left( 1+\frac{\alpha}{2}\,\phi^{\prime}\,^{2} \right)%
    \phi^{\prime}\,^{2}-\frac{2}{3}\,V\,,
\end{equation}
\begin{equation}\label{em3b}
    A^{\prime}\,^{2}=\frac{1}{6}\left( 1+\frac{3}{2}\,\alpha \phi^{\prime}\,^{2} \right)\phi^{\prime}\,^{2}
    -\frac{1}{3}\,V\,.
\end{equation}
\noindent It is possible to rewrite~(\ref{em2b}) and~(\ref{em3b}) as
\begin{equation}\label{em3c}
A^{\prime\prime}=-\frac{2}{3}\,\phi^{\prime}\,^{2}\left(1+\alpha \phi^{\prime}\,^{2}\right)\,.
\end{equation}
Now, to extend the first-order framework to the braneworld scenario, we follow
the work in~\cite{sken}, and choose the derivative of the warp factor with respect to the
extra dimension to be a function of the scalar field
\begin{equation}\label{a}
    A^{\prime}=-\frac{1}{3}\,W(\phi).
\end{equation}

\begin{figure}[ht]
\begin{center}
\includegraphics[width=6cm, angle=0]{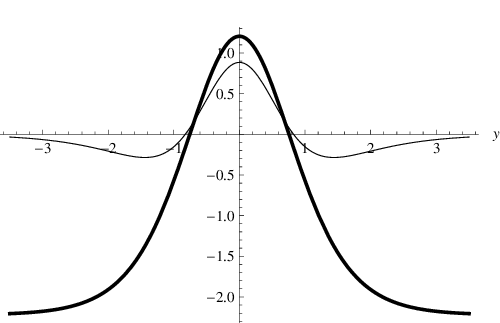}
\end{center}
\caption{The profiles of the matter energy density (thin line) and Ricci scalar %
         (thick line) for $b=2/3$, $c=1$ and $\alpha=0.1$.} \label{desc}
\end{figure}

\noindent Substituting~(\ref{a}) into~(\ref{em3c}), we get
\begin{equation}\label{em4b}
    \phi^{\prime}+\alpha\phi^{\prime}\,^{3}=\frac{1}{2}\,W_{\phi}\,,
\end{equation}
\noindent this equation is a cubic equation in $\phi^{\prime}$, then
the real solution to this cubic equation is given by
\begin{equation}\label{phi1}
    \phi^{\prime}=\frac{m(W_{\phi})}{6\alpha}-\frac{2}{m(W_{\phi})}\,,
\end{equation}
\noindent where
\begin{equation}\label{mphi}
    m(W_{\phi})=\left(54\alpha^{2}W_{\phi}+6\sqrt{3}\left( 16\alpha^{3}+ %
    27\alpha^{4}W_{\phi}^{2} \right)^{1/2}  \right)^{1/3}.
\end{equation}
\noindent It is instructive to note that the equation~(\ref{phi1}) is the first-order
differential equation for the scalar field $\phi$.
\noindent The potential is obtained by substituting~(\ref{a}) in~(\ref{em3b})
\begin{equation}\label{v}
    V(\phi)=\frac{1}{2}\,\phi^{\prime}\,^{2}+ \frac{3}{4}\,\alpha\phi^{\prime}\,^{4}-
    \frac{1}{3}W(\phi)^{2}\,,
\end{equation}
\noindent where $\phi^{\prime}$ is given by~(\ref{phi1}).
On the other hand, we consider the energy functional~\cite{sken}
\begin{equation}\label{ef}
    E[A,\phi]=\int dy(-\mathcal{L}_{system}),
\end{equation}
\noindent where $\mathcal{L}_{system}=\sqrt{|g|}\left[-R/4+\mathcal{L}(\phi,X)\right]$. This energy functional coincides with the Hamiltonian deduced from Einstein-Hilbert action (including a surface term), which is well-defined for spatially noncompact geometries and it is fully agree with the definition of the total energy in general relativity, as was reported by
Hawking~\cite{cqg}. In our case,~(\ref{ef}) becomes
\begin{equation}\label{es2}
E\left[ A,\phi \right]=\int_{-\infty}^{\infty} dy \,\mathrm{e}^{4A}\left\{ \frac{1}{2}\,%
\phi^{\prime}\,^{2}-3A^{\prime}\,^{2}+\frac{\alpha}{4}\,\phi^{\prime}\,^{4} +V\right\}.
\end{equation}
\noindent The Euler-Lagrange differential equations from the functional~(\ref{es2}) are
the static equations of motion~(\ref{em1b}) and~(\ref{em2b}), this result is valid for
all $\alpha$. Otherwise, substituting the potential~(\ref{v}) in~(\ref{es2}), for $\alpha$
which is not necessarily small, we can not deduce the first order differential equations%
~(\ref{a}) and~(\ref{phi1}) from the energy functional, therefore the solutions of the
first order differential equations not necessarily minimize the energy functional.
\noindent At this point, it is also instructive to analyze the matter energy in the model
with nonstandard dynamics. From~(\ref{de}),~(\ref{dlg}) and~(\ref{v}), we get
\begin{equation}\label{rhog2}
    E_{\phi}=\int_{-\infty}^{\infty} dr \,\mathrm{e}^{2A}\left\{ \phi^{\prime}\,^{2}+%
    \alpha\phi^{\prime}\,^{4} - \frac{1}{3}\,W^{2}\right\}\,.
\end{equation}
\noindent Finally, using~(\ref{em3c}) and~(\ref{a}) and integrating, we obtain
\begin{equation}\label{rhog3}
    E_{\phi}=\frac{1}{2}\,\left(\mathrm{e}^{2A(\infty)}W(\phi(\infty))-%
    \mathrm{e}^{2A(-\infty)}W(\phi(-\infty))\right)\,,
\end{equation}
\noindent the value of the matter energy for all $\alpha$. Note that the matter
energy depends on the asymptotic behavior of the warp factor.
Now, we follow the same procedure of~\cite{ba3} and let us focus our study
in the case of $\alpha$ very small. Thus, the solution of~(\ref{em4b}), up to first-order in $\alpha$ becomes
\begin{equation}\label{phi2}
    \phi^{\prime}=\frac{1}{2}\,W_{\phi}-\frac{\alpha}{8}\,W_{\phi}^{3}\,,
\end{equation}
\noindent and substituting~(\ref{phi2}) into~(\ref{v}) we obtain the potential
\begin{equation}\label{v2}
    V(\phi)=\frac{1}{8}\,W_{\phi}^{2}-\frac{\alpha}{64}\,W_{\phi}^{4}- \frac{1}{3}\,W^{2}\,.
\end{equation}
At this point we turn to examine the energy functional~(\ref{es2}). Substituting the potential~(\ref{v2}) in~(\ref{es2}), we get
$$E\left[A,\phi\right]=\int_{-\infty}^{\infty}dy\left\{\frac{1}{2}\left(\phi^{\prime}-\frac{1}{2}W_{\phi}+%
\frac{\alpha}{8}W_{\phi}^{3}\right)^{2}\right. $$
$$\hspace{0.35in}\left.-3\left(A^{\prime}+\frac{1}{3}W\right)^{2}\right\}+\frac{\alpha}{8}\int_{-\infty}^{\infty}dy\left(2\phi^{\prime 4}+\frac{3}{8}W_{\phi}^{4}\right.
$$
\begin{equation}\label{5.51}
\hspace{0.4in}\left.-W_{\phi}^{3}\phi^{\prime}\right)+\frac{1}{2}\int_{-\infty}^{\infty}dy\frac{d}{dy}\left(We^{4A}\right).
\end{equation}
\noindent From~(\ref{5.51}) we get a important new result, the solutions of the first order
differential equations,~(\ref{a}) and~(\ref{phi2}), are those that minimize the
energy functional. Thus, the first order differential equations can
be seen as the BPS equations and the energy functional could play
the role of a BPS energy in such scenario. The value of the energy system
for $\alpha$ very small is given by
\begin{equation}
E\left[A,\phi\right]=\left|e^{4A(\infty)}W(\phi(\infty))-e^{4A(-\infty)}W(\phi(-\infty))\right|.
\label{eq5.52}
\end{equation}
\noindent Again, the asymptotic behavior of the warp factor plays a
leading role in the value of the energy functional.
Now, we find explicitly the solutions for~(\ref{a}) and~(\ref{phi2}), which
minimizes the energy functional. Then, the solution for~(\ref{phi2}) becomes
\begin{equation}\label{phi3}
    \phi(y)=\phi_{0}(y)-\frac{\alpha}{4}\,W_{\phi}(\phi_{0}(y))W(\phi_{0}(y))\,,
\end{equation}
\noindent where $\phi_{0}(y)$ is the solution when $\alpha=0$. From~(\ref{a}) and%
~(\ref{phi3}), we obtain
\begin{equation}\label{a1}
    A(y)=A_{0}(y)+\frac{\alpha}{12}\,W(\phi_{0}(y))^{2},
\end{equation}
\noindent where $A_{0}(y)$ represents $A(y)$ when $\alpha=0$. The matter energy density
given by~(\ref{de}) is
\begin{equation}\label{de2}
    \rho=\mathrm{e}^{2A(y)}\left( \frac{1}{4}\,W_{\phi}^{2}-\frac{1}{3}\,W^{2}
    -\frac{\alpha}{16}\,W_{\phi}^{4} \right),
\end{equation}
\noindent substituting~(\ref{phi3}) and~(\ref{a1}) in~(\ref{de2}), we obtain
\begin{equation}\label{de3}
    \rho=\rho_{0}-\frac{\alpha}{48}\,\mathrm{e}^{2A_{0}(y)}G_{0}\,,
\end{equation}
\noindent where
\begin{equation}\label{rho0}
    \rho_{0}=\mathrm{e}^{2A_{0}(y)}\left( \frac{1}{4}\,W_{\phi}^{2}-\frac{1}{3}\,
    W^{2} \right)_{\phi=\phi_{0}}\,
\end{equation}
\noindent and
\begin{equation}\label{G0}
G_{0}=\left( 6W_{\phi\phi}W_{\phi}^{2}W-%
    10W^{2}W_{\phi}^{2}+3W_{\phi}^{4}+\frac{8}{3}\,W^{4}\right)_{\phi=\phi_{0}}
\end{equation}
\noindent note that the energy density~(\ref{de3}) is a little bit different
from that given in~\cite{ba3}.
To show the validity of the solutions,~(\ref{phi3}),~(\ref{a1}) and~(\ref{de3}),
we consider the superpotential $W(\phi)$ of the form~\cite{grem}
\begin{equation}\label{sp1}
    W(\phi)=3\,b\, c\sin\left(\sqrt{\frac{2}{3b}}\,\phi\right)\,,
\end{equation}
\noindent where $b$ and $c$ are positive parameters. The classical solutions for~(\ref{phi3})
and~(\ref{a1}) are given by
$$\phi(y)=\sqrt{\frac{3b}{2}}\arcsin\left[ \tanh\left(c\,y\right) \right]-\frac{3\sqrt{6}\alpha}{4}\,b^{3/2}c^{2} $$
\begin{equation}\label{sol}
\hspace{0.8in}\times\tanh\left(c\,y\right) \mathrm{sech}\left(c\,y\right) ,
\end{equation}
\noindent and
\begin{equation}\label{ay}
    A(y)=b\ln\left[ \mathrm{sech}\left( c\,y \right) \right]+\frac{3\alpha}{4}\,b^{2}c^{2}\tanh^{2}\left(c\,y\right).
\end{equation}
\noindent The profiles of the matter energy density and the Ricci scalar
are shown in fig.~\ref{desc} for $\alpha=0.1$. A similar behavior is
obtained for $\alpha=1$ and $\alpha= 10$. Note that the presence of regions
with positive Ricci scalar is connected with the localization of the brane. Also
note that far from the brane, $R$ tends to a negative constant, characterizing the
$AdS_{5}$ limit from the bulk. In the limit $|y|\rightarrow\infty$, the warp factor is $\mathrm{e}^{2A}\rightarrow0$, therefore the matter energy~(\ref{rhog3}) and the energy of system~(\ref{eq5.52}) both are zero and the thick brane solution coincides with the solution of RS model as $bc=constant$~\cite{rome}.

Now on, we can analyze the exact solutions in the thin brane limit ($c\rightarrow\infty$ and the product $bc$ is held fixed)~\cite{rome,mann}, we obtain
\begin{equation}\label{tb1}
    \phi(y)=\frac{\sqrt{6b}}{4}\,\pi \,\mathrm{sgn}(y)\,,
\end{equation}
\begin{equation}\label{tb2}
    A(y)=-bc|\,y|+\frac{3\alpha}{4}\,b^{2}c^{2}\,,
\end{equation}
\noindent where the term $\frac{3\alpha}{4}\,b^{2}c^{2}$ in~(\ref{tb2}) is the contribution of nonstandard kinetic term on the branemodel geometry. From this we get another new result, the solutions of the braneworld model treated here is consistent, because when $\alpha=0$ it reduces to the solutions of RS model (thin brane model).

\section{Fermion localization}

The action for a Dirac spinor field coupled with the scalar fields by a general Yukawa coupling is
\begin{equation}\label{ad}
    S=\int d^{5}x\sqrt{|\,g|}\left[ i\bar{\Psi}\Gamma^{M}\nabla_{M}\Psi-\eta\bar{\Psi}F(\phi)\Psi \right]\,,
\end{equation}
\noindent where $\eta$ is the positive coupling constant between fermions and the scalar field. Moreover,
we are considering the covariant derivative $\nabla_{M}=\partial_{M}+\frac{1}{4}\,\omega^{\bar{A}\bar{B}}_{M}\Gamma_{\bar{A}}\Gamma_{\bar{B}}$\,,
where $\bar{A}$ and $\bar{B}$, denote the local Lorentz indices and $\omega^{\bar{A}\bar{B}}_{M}$ is
the spin connection. Here we consider the field $\phi$ as a background field. The equation of motion is
obtained as
\begin{equation}\label{dkp}
i\,\Gamma ^{M }\nabla_{M}\Psi-\eta F(\phi)\Psi =0.
\end{equation}%
\noindent We choose the irreducible representation $\Gamma^{\mu}=e^{-A}\gamma^{\mu}$ and $\Gamma^{4}=-i\gamma^{5}$. The equation of motion~(\ref{dkp}) becomes
\begin{equation}\label{em}
\left[ i\gamma^{\mu}\partial_{\mu}+\gamma^{5}\mathrm{e}^{A}(\partial_{y}+2\partial_{y}A)%
-\eta\,\mathrm{e}^{A}F(\phi) \right]\Psi=0.
\end{equation}
\noindent Now, we use the general chiral decomposition
\begin{equation}\label{dchiral}
    \Psi(x,y)=\sum_{n}\psi_{L_{n}}(x)\alpha_{L_{n}}(y)+\sum_{n}\psi_{R_{n}}(x)\alpha_{R_{n}}(y),
\end{equation}
\noindent with $\psi_{L_{n}}(x)=-\gamma^{5}\psi_{L_{n}}(x)$ and $\psi_{R_{n}}(x)=\gamma^{5}\psi_{R_{n}}(x)$.
With this decomposition $\psi_{L_{n}}(x)$ and $\psi_{R_{n}}(x)$ are the left-handed and
right-handed components of the four-dimensional spinor field, respectively. After applying
~(\ref{dchiral}) in~(\ref{em}), and demanding that $i\gamma^{\mu}\partial_{\mu}\psi_{L_{n}}=m_{n}\psi_{R_{n}}$
and $i\gamma^{\mu}\partial_{\mu}\psi_{R_{n}}=m_{n}\psi_{L_{n}}$, we obtain two equations
for $\alpha_{L_{n}}$ and $\alpha_{R_{n}}$
\begin{equation}\label{ea1}
    \left[ \partial_{y}+2\partial_{y}A+\eta F(\phi) \right]\alpha_{L_{n}}=m_{n}\mathrm{e}^{-A}\alpha_{R_{n}}\,,
\end{equation}
\begin{equation}\label{ea2}
    \left[ \partial_{y}+2\partial_{y}A-\eta F(\phi) \right]\alpha_{R_{n}}=-m_{n}\mathrm{e}^{-A}\alpha_{L_{n}}\,.
\end{equation}
\noindent The orthonormality condition for $\alpha_{L_{n}}$ and $\alpha_{R_{n}}$ is given by
\begin{equation}\label{orto}
    \int_{-\infty}^{\infty}dy\,\mathrm{e}^{3A}\alpha_{Lm}\alpha_{Rn}=\delta_{LR}\delta_{mn}.
\end{equation}
\noindent Implementing the change of variables %
\begin{equation}\label{cv}
    z=\int^{y}_{0}\mathrm{e}^{-A(y^{\,\prime})}dy^{\,\prime},
\end{equation}
\noindent $\alpha_{L_{n}}=\mathrm{e}^{-2A}L_{n}$ and $\alpha_{R_{n}}=\mathrm{e}^{-2A}R_{n}$, we get
\begin{equation}\label{sleft}
    -L_{n}^{\prime\prime}(z)+V_{L}(z)L_{n}=m_{n}^{2}L_{n}\,,
\end{equation}
\begin{equation}\label{sright}
    -R_{n}^{\prime\prime}(z)+V_{L}(z)R_{n}=m_{n}^{2}R_{n}\,,
\end{equation}
\noindent where
\begin{eqnarray}
  V_{L}(z) &=& \eta^{2}\mathrm{e}^{2A}F^{2}(\phi)-\eta\partial_{z}\left( \mathrm{e}^{A}F(\phi) \right),\label{vefa} \\
  V_{R}(z) &=& \eta^{2}\mathrm{e}^{2A}F^{2}(\phi)+\eta\partial_{z}\left( \mathrm{e}^{A}F(\phi) \right)\label{vefb}.
\end{eqnarray}
\noindent Using the expressions $\partial_{z}A=\mathrm{e}^{A(y)}\partial_{y}A$ and
$\partial_{z}F=\mathrm{e}^{A(y)}\partial_{y}F$, we can recast the potentials~(\ref{vefa}) and~(\ref{vefb})
as a function of $y$~\cite{yo}
\begin{eqnarray}
  V_{L}(z(y)) &=& \eta\mathrm{e}^{2A}\left[ \eta F^{2}-\partial_{y}F-F\partial_{y}A(y) \right],\label{vya} \\
  V_{R}(z(y)) &=& V_{L}(z(y))|_{\eta\rightarrow-\eta}\,.\label{vyb}
\end{eqnarray}
\noindent Now we focus attention on the calculation of the zero mode. Substituting $m_{n}=0$ in~(\ref{ea1}) and~(\ref{ea2})
and using $\alpha_{L_{n}}=\mathrm{e}^{-2A}L_{n}$ and $\alpha_{R_{n}}=\mathrm{e}^{-2A}R_{n}$, respectively, we get
\begin{equation}\label{mzL}
    L_{0}\propto \exp \left[-\eta\int_{0}^{y}dy^{\prime}F(\phi) \right],
\end{equation}
\begin{equation}\label{mzR}
    R_{0}\propto \exp \left[\eta\int_{0}^{y}dy^{\prime}F(\phi) \right].
\end{equation}
\noindent This fact is the same to the case of two-dimensional Dirac equation (isolated solutions)~\cite{luis}. In order to guarantee the normalization condition~(\ref{orto}) for the left-handed fermion zero
mode~(\ref{mzL}), the integral must be convergent, \textit{i.e}
\begin{equation}\label{cono}
    \int^{\infty}_{-\infty}dy\exp\left[ -A(y)-2\eta\int^{y}_{0}dy\,^{\prime}F(\phi(y\,^{\prime})) \right]<\infty.
\end{equation}
\noindent This result clearly shows that the behavior of $F(\phi(y))$ plays a leading role
for the fermion localization on the brane~\cite{yo}.
\begin{figure}[ht]
\begin{center}
\includegraphics[width=7 cm, angle=0]{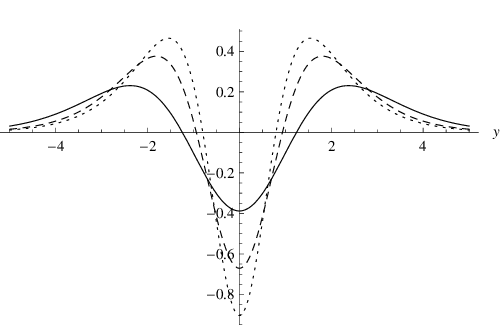}
\end{center}
\par
\vspace*{-0.1cm} \caption{Potential profile of $V_{L}(y)$ for $\eta=1$, $b=2/3$,  $c=1$, $\alpha=0.1$ (dotted line), $\alpha=1$ (dashed line) and $\alpha=10$ (thin line).} \label{pea}
\end{figure}
\begin{figure}[ht]
\begin{center}
\includegraphics[width=7 cm, angle=0]{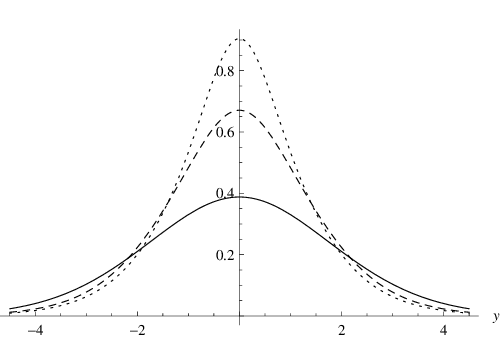}
\end{center}
\par
\vspace*{-0.1cm} \caption{Potential profile of $V_{R}(y)$ for $\eta=1$, $b=2/3$,  $c=1$, $\alpha=0.1$ (dotted line), $\alpha=1$ (dashed line) and $\alpha=10$ (thin line).} \label{peb}
\end{figure}

\subsection{Zero mode and fermion localization}
From now on, we mainly consider the simplest case $F(\phi)=\phi$. First, we consider the
normalizable problem of the solution. In this case, from eqs.~(\ref{sol}) and~(\ref{ay})
the integrand in (\ref{cono}) can be expressed as
$$I=\exp\left[ -b\ln\left( \mathrm{sech}\left( c\,y \right) \right)-\frac{3\alpha }{4}\,%
b^{2}c^{2}\tanh^{2}\left(c\,y\right)\right.$$
\begin{equation}\label{int1}
\hspace{0.5in}\left.-\eta\sqrt{6b}\, \bar{I}(y)-\eta\frac{3\sqrt{6}\alpha}{2}\,b^{3/2}c\,\mathrm{sech}(cy)\right],
\end{equation}
\noindent where $\bar{I}=\int dy^{\prime}\arcsin\left[ \tanh(cy^{\prime})\right]$. Following
the same procedure of~\cite{mel}, we only need to consider the asymptotic behavior of
the integrand. It becomes
\begin{equation}\label{in}
    I\rightarrow\mathrm{exp}\left[ -\left( \eta\,\pi\sqrt{\frac{3b}{2}}-bc \right)|\,y| -\frac{3}{4}\,\alpha b^{2}c^{2}\right]\,.
\end{equation}
\noindent This result clearly shows that if $\eta>\frac{c}{\pi}\,\sqrt{\frac{2b}{3}}$ the zero mode of the left-handed fermions is normalized. Note that the asymptotic behavior of the normalization condition for this case is independent of $\alpha$. For the right-handed fermions, we can use the change $\eta\rightarrow-\eta$ (that implies $L_{0}\rightarrow R_{0}$) in (\ref{in}) and we can conclude that the right-handed fermions can not be a normalizable zero mode. The shape of the potentials for this case are shown in figures~\ref{pea} and~\ref{peb} for some values of $\alpha$. The Fig.~\ref{pea} shows that the potential of left-handed fermions, $V_{L}$, is indeed a volcano-like potential and that the depth of the well structure decreases as $\alpha$ increases. From this, we can conclude that the ability to trap fermions of the effective potential $V_{L}$ is inversely proportional to $\alpha$. On the other hand, Figure~\ref{peb} shows that the potential $V_{R}$ is always positive (none bound fermions) and this effective potential has a maximum that decreases as $\alpha$ increases. The analytic expressions for this model are only valid for $\alpha$ small. For $\alpha$, which is not necessarily small, the numerical study is used. The numerical study done for a large range of values of $\alpha$ bear out our results.

\section{Conclusions}
We have considered the braneworld model with nonstandard kinetic terms $\mathcal{L}=K(X)-V(\phi)$, where
$K=X+\alpha|\,X|X$ (type I model in~\cite{ba3}). This letter completes the analysis of the meritorious research in~\cite{ba3}. We showed that the equations of motion for all $\alpha$ can be deduced
from the functional $E[A,\phi]$~(\ref{es2}), as done by Townsend~\cite{sken} in the case of standard dynamics.
Furthermore, we showed that for $\alpha$ small the solutions of the first order differential equations, %
~(\ref{a}) and~(\ref{phi2}), are those that minimize the energy of system. In contrast, for $\alpha$,
which is not necessarily small, the solutions of the first order differential equations not necessarily
minimize the energy of system. Also, we showed that the value of the matter energy and the energy of
system depends on the asymptotic behavior of the warp factor. We found an expression for the matter
energy density that differs slightly from~\cite{ba3}. The numerical study gives full support to our
matter energy density expression for $\alpha$ small. Furthermore, we showed that the braneworld model
with nonstandard dynamics treated here is consistent, because it reduces to the RS model
(thin brane model) for $\alpha=0$.
We also have investigated the localization problem of fermions for the type I model. We have used
the simplest Yukawa coupling $\eta\bar{\Psi}\phi\Psi$ between the scalar and the spinor fields. In order
to guarantee the normalization condition for the zero mode, we showed that the zero mode of left-handed
fermions is normalizable under the condition $\eta>\frac{c}{\pi}\,\sqrt{\frac{2b}{3}}$ and it is
independent of $\alpha$. For this kind of solution, the effective potential of left-handed fermions
$V_{L}$ is a volcano-like potential. $V_{L}$ has a minimum at the localization of the brane ($y=0$),
therefore the zero mode of the left-handed is localized on the brane. On the other hand, the value of $\alpha$ adjust the minimum of $V_{L}$, the depth of the well structure decreases as $\alpha$ increases. Therefore, we can conclude that the ability to trap fermions of $V_{L}$ is inversely proportional to
$\alpha$. For $\alpha$ not necessarily small, the numerical study done for a large range of values of $\alpha$ bear out our results and conclusions. Finally, based on our results, we intend to expand the present study to analyze the in\-te\-res\-ting problem of localization of another spin fields.

\acknowledgments
The authors would like to thank Professors J. M. Hoff da Silva and M. B. Hott for useful discussion. The authors are also indebted to the anonymous referee for an excellent and constructive review. This work was supported in part by means of funds provided by CAPES. L.E.A.M. thanks the Brazilian funding agency CAPES for support through a scholarship under the PEC-PG program.

\end{document}